\documentclass[
  aps,
  prb,
  reprint,
  superscriptaddress,
  amsmath,
  amssymb,
]{revtex4-2}

\usepackage{graphicx}
\usepackage{dcolumn}
\usepackage{bm}
\usepackage{siunitx}
\usepackage[
    hidelinks
]{hyperref}

\begin{document}
\preprint{APS/123-QED}

\title{Magnetotransport experiments on side-contacted topological insulator three-terminal Josephson junctions}

\author{Christoph Wassmer}
\author{Ella Nikodem}
\author{Oliver Breunig}
\author{Yongjian Wang}
\altaffiliation{Present address: High Magnetic Field Laboratory of Anhui Province, HFIPS, Chinese Academy of Sciences, Hefei 230031, China}
\author{Yoichi Ando}
\author{Jakob Schluck}

\email{schluck@ph2.uni-koeln.de}
\affiliation{
 Physics Institute II, University of Cologne, Zülpicher Straße 77, 50937 Köln, Germany
}
\date{\today}


\begin{abstract}

Multiterminal Josephson junctions have recently attracted substantial attention. The ability to study Andreev bound states in a multi-phase parameter space opens up exciting avenues for generating topologically non-trivial phases in superconducting devices.
A particularly interesting case is the trijunction based on a topological insulator (TI), which is predicted to host Majorana zero modes for suitable phase configurations (Fu and Kane, PRL 2008).
Here we turn to the platform of the side-contacted TI trijunction and present experimental evidence for a multiterminal Josephson effect.
We study the transport properties and find typical features of a two-dimensional critical current contour (CCC) as well as bias regimes in which only a single junction remains dissipationless.
We pay particular attention to these features and study the effects of magnetic fields on single junctions within the trijunction, where, depending on the field direction, the supercurrents show Fraunhofer patterns, SQUID oscillations, or a monotonic decay.
We find good agreement with semiclassical simulations for a single TI junction that take into account the top and bottom surface states.
We observe pronounced non-convex shape deformations of the CCC under perpendicular magnetic fields that reflect the flux buildup in the loop formed by our trijunctions and provide a simple model sufficient to explain the observed effect.
Our results provide the foundations for future studies of Majorana zero modes on this platform.  

\end{abstract}

\maketitle


\section{Introduction}
Two superconductors, coupled via a weak link, form a Josephson junction \cite{Josephson1962}, a device that can carry a dissipationless supercurrent.
The critical current is governed by the superconducting phase difference between the two superconductors.
The Josephson junction has been studied extensively for decades, and devices based on it form the basis of modern superconducting electronics \cite{Barone1982}. They have applications, among others, in SQUID magnetometry \cite{Fagaly2006}, frequency mixers \cite{Tucker1985}, quantum computing \cite{Arute2019}, and superconducting diodes \cite{Baumgartner2022}.

A natural generalization of a Josephson junction emerges by coupling multiple superconducting electrodes via the same weak link.
In such a system, the relation between the individual superconducting phases and the supercurrent is complex.
When the weak link is a normal metal, the Andreev bound states (ABS) formed in such a system \cite{Golubov2004} effectively realize a multi-dimensional band structure.
It was realized that, theoretically, this would allow for the generation of topological phases in otherwise non-topological materials \cite{Riwar2016}.

Experimentally, recent progress in control over the superconducting proximity effect in semiconductor materials \cite{Chang2015} and the availability of hexagonal boron nitride, enabling quality-preserving fabrication of two-dimensional materials \cite{Dean2010}, allowed for the realization of transport studies in multiterminal Josephson junctions and the identification of characteristic signatures \cite{Draelos2019,Pankratova2020,Graziano2020}. Both the critical current contour (CCC), the multi-dimensional generalization of the critical current, and the situation, where a single junction is dissipationless while the others are driven to a resistive state were studied. This results in characteristic resistance drops along star-like lines originating from the CCC in the higher-dimensional bias space. An exciting possibility is the observation of signatures of higher-order non-local Cooper pair transport on this platform \cite{Cohen2018,Huang2022}, but caution has been raised that semiclassical network effects can mimic such signatures \cite{Zhang2023}. 
More recent experiments focus on the tunability of the ABSs directly by studying their spectroscopic signatures \cite{Coraiola2024a,Coraiola2024b,Antonelli2025}.
In this regard, a material class of particular interest is topological insulators (TIs) \cite{Ando2013}. It was realized early on by Fu and Kane \cite{Fu2008} that trijunctions made of TIs proximity-coupled to ordinary superconductors can be employed to generate a discrete vortex in the system, which hosts an isolated Majorana zero mode (MZM). These are of great interest because of their non-Abelian statistics and possible applications in topological quantum computing \cite{Nayak2008,Sato2017}. Despite tremendous efforts and encouraging results \cite{Yazdani2023}, to date these MZMs remain elusive.
Top-contacted TI trijunctions have been studied in transport \cite{Koelzer2023,Behner2025} and also by contact resistance spectroscopy \cite{Yang2019,Zhang2022}, but so far clear signatures of MZMs are missing.
Here we turn to the platform of side-contacted three-dimensional TI Josephson junctions, recently introduced by us \cite{Nikodem2025,Nikodem2025diode,Nikodem2026}. This geometry allows for a robust proximity effect on all TI surfaces, reducing the problem of excess normal carriers and hybridized end-states.
We demonstrate the fabrication of high-quality trijunctions and study their response to external magnetic fields applied in-plane along different directions as well as out-of-plane.
We find that the effect of an in-plane field can be well understood by a semiclassical trajectory model for the individual junctions and, depending on the field direction, results in the known nano-SQUID oscillations \cite{Nikodem2025}  or the suppression of the critical current \cite{Chen2018}.
By considering the fluxoid quantization in the loop formed by the trijunction, the out-of-plane response can be reproduced reasonably well within an RCSJ model.
Finally, we study the effect of an applied back-gate voltage and find an unexpected increase in the critical current of one of the junctions.
The paper is structured as follows. First, we show the device fabrication and demonstrate the typical trijunction features in the multi-dimensional current-bias space.
We proceed by isolating the response of the individual junctions to an external magnetic field along different device directions before turning to the two-dimensional characterization.
After presenting the effect of an applied gate voltage, the results are discussed. 


\section{Side-contacted topological-insulator Three-Terminal Josephson Junction}

\begin{figure}[t]
\includegraphics[width= 3.35in]{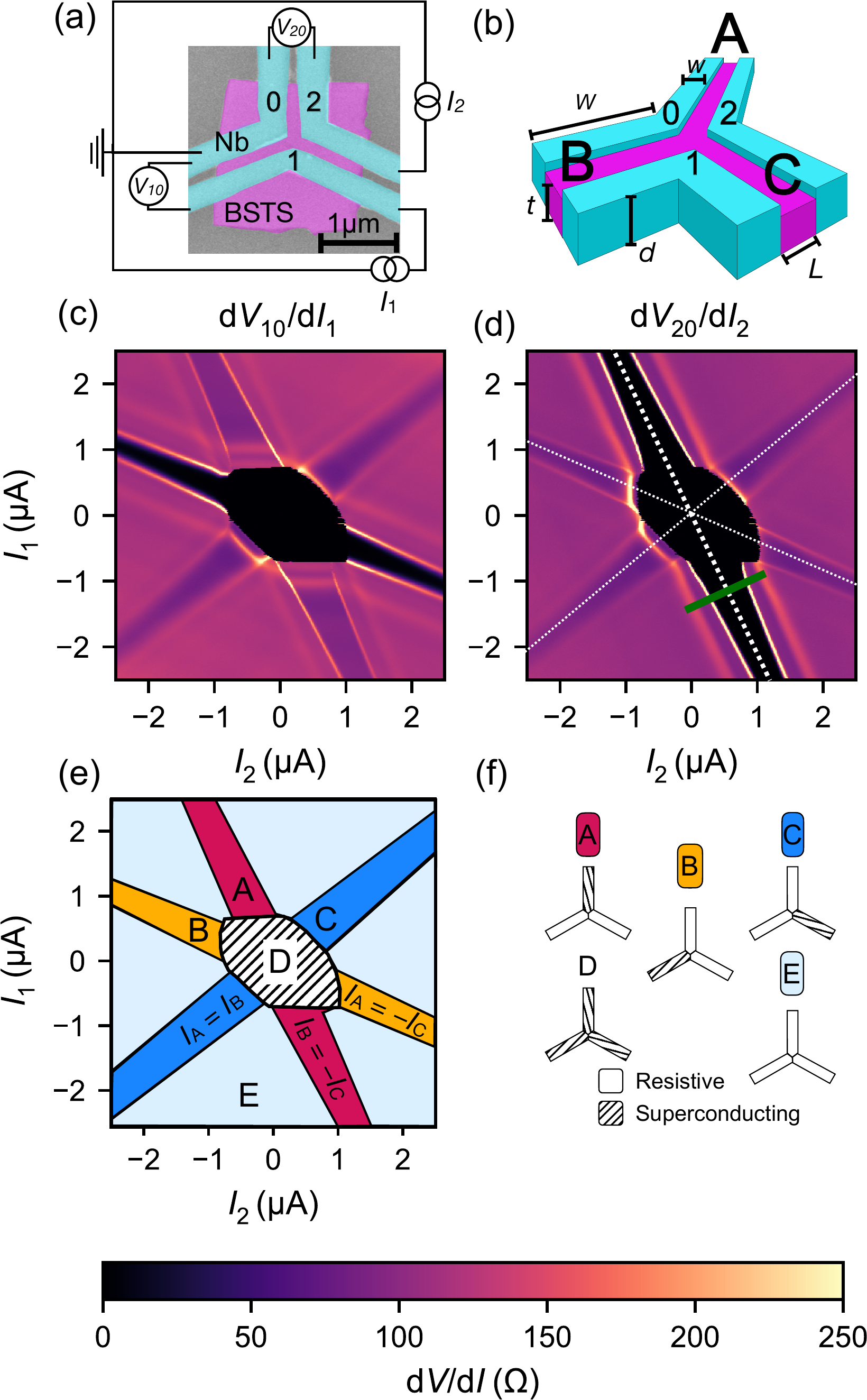}
\caption{
    \label{fig:1} 
    Side-contacted BSTS-Nb trijunction.
    \textbf{(a)} SEM image and measurement configuration. Terminals 1 and 2 are current-biased while terminal 0 is electrically grounded. Si substrate with \SI{300}{nm} SiO$_2$, Si back gate not shown in sketch.
    \textbf{(b)} Schematic 3D view, three superconducting Nb-leads (turquoise) of height $d$ are connected by a central TI island of thickness $t$ (pink).
    \textbf{(c)} Bias map showing d$V_{10}$/d$I_{1}$ in bias space $I_{1,2}$.
    \textbf{(d)} Bias map showing d$V_{20}$/d$I_{2}$ in bias space $I_{1,2}$ including visual guides.
    White dashed lines indicate bias configurations for which the respective junction carries zero current.
    Measurement cut used in Sec. \ref{sec:singlefield} highlighted in dark green.
    \textbf{(e,f)} Illustration of the different transport regimes observed in the trijunction. 
    {A}, {B}, and {C} refer to the junction that remains in the zero-resistance state, {D} refers to the critical current contour, {E} refers to the generic resistive state.
        }
\end{figure}

Our devices consist of three superconducting leads connected via a single Y-shaped topological-insulator (TI) nanowire.
A false-colored SEM image together with a schematic illustration of the device is shown in Fig.~\ref{fig:1} (a,b).
First, an exfoliated $\mathrm{BiSbTeSe_2}$(BSTS)\cite{Ren2011} flake is etched into a planar Y-shaped geometry by an Ar-ion beam.
Nb leads are then deposited in situ via dc-sputtering along the sidewalls of the nanowire, forming three lateral junctions. 
In the bulk-insulating TI, transport is confined to the top and bottom surfaces of the central nanowire \cite{Nikodem2025}. 
The resulting trijunction consists of three terminals, labeled 0, 1, and 2, interconnected by three SNS-type Josephson junctions denoted A, B, and C of comparable length and width.
The device geometry is defined by inter-terminal angles of \SI{120}{\degree}, a Nb lead width of $w = \SI{400}{nm}$, and three junctions with a length of $L \approx \SI{140}{nm}$ and a width of $W \approx$ \SIrange{0.8}{1}{\micro\meter}.
The nanowire has a thickness of $t \approx \SI{50}{nm}$ and a width of \SI{140}{nm}, corresponding to the junction length. 
The measurement configuration is shown in Fig.~\ref{fig:1}(a).
Two independent DC bias currents, $I_1$ and $I_2$, are applied via terminals 1 and 2, while the differential resistances d$V_{10}$/d$I_{1}$ and d$V_{20}$/d$I_{2}$ are measured with respect to the grounded terminal 0 using standard lock-in techniques.
All measurements are performed in a $^{3}$He cryostat with a base temperature of $T \approx \SI{260}{mK}$.

Fig.~\ref{fig:1}(c) shows the differential resistance between terminals 1 and 0, while Fig.~\ref{fig:1}(d) displays the corresponding map for terminals 2 and 0, including visual guides.
Five distinct transport regimes, labeled A-E, emerge depending on the bias-current configuration, resulting in three types of characteristic features \cite{Pankratova2020,Graziano2020}. For clarity, we schematically show the different regimes in Fig.~\ref{fig:1}(e) and the underlying state of the single junctions in Fig.~\ref{fig:1}(f).
The superconducting state of the three-terminal device is described by three phase variables $\varphi_\mathrm{A,B,C}$, with only two independent phase differences.
As a result, the critical current is no longer characterized by a single scalar quantity but by a boundary in the two-dimensional $(I_1, I_2)$ current space.
For sufficiently small bias currents, all three junctions stay in the zero-resistance state marked {D} in the center region of Fig.~\ref{fig:1}(e), also referred to as the CCC.

For bias combinations resulting in $V_i = 0$, where $V_i$ denotes the voltage drop across junction $i$ ($i \in \{\mathrm{A,B,C}\}$), arm-like features of reduced d$V$/d$I$ appear along the white dashed lines in Fig.~\ref{fig:1}(d).
This drop in differential resistance can be understood within a simple resistor-network picture as discussed in previous work \cite{Graziano2020}.
Along these lines, the bias configuration effectively suppresses the net current through junction $i$, as current partitioning in the resistor-network model is set by Kirchhoff’s laws and the relative junction resistances $R_{\rm A}$, $R_{\rm B}$, and $R_{\rm C}$.
Current flows only across the other two junctions, while junction $i$ experiences zero bias current and thus remains in the zero-resistance state.
The width of the arms is set by the corresponding critical currents and is further reduced by Joule heating.
Region {E} denotes the regime in which all three junctions are in the resistive state. The transport features observed in this region are attributed to multiple Andreev reflection (MAR) processes in the junctions.
We have thereby demonstrated that our device indeed forms a trijunction showing all the typical transport features~\cite{Pankratova2020,Graziano2020,Behner2025}. However, signatures of multiplets are absent in our device.


\section{Isolating the Magnetic-Field Response of individual Junctions}
\label{sec:singlefield}

\begin{figure}[t]
\includegraphics[width= 3.35in]{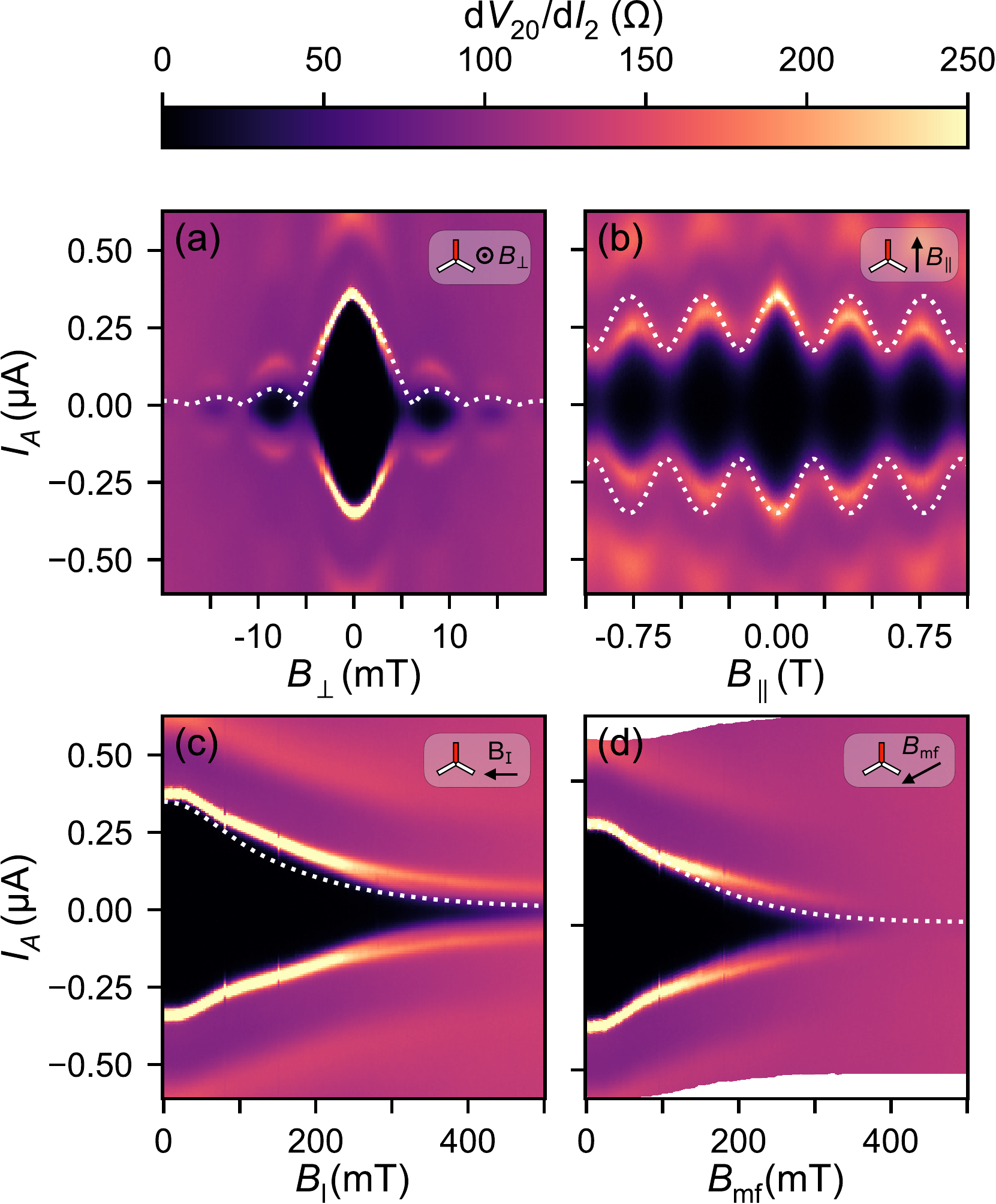}
\caption{
    \label{fig:2}
    Magnetic-field response of isolated junction A.
    Data are measured along the dark green line shown in Fig.~\ref{fig:1}(d), insets indicate the applied field direction and highlight the measured junction (A) in red. The white dashed lines show the result of a semiclassical calculation of the critical current for a single junction in magnetic field.
    \textbf{(a)} Fraunhofer pattern in d$V$/d$I$ as a function of out-of-plane field. The field is always swept starting from zero.
    \textbf{(b)} SQUID pattern in d$V$/d$I$ as a function of in-plane field applied \textit{along} junction A, \textit{across} direction of current flow.
     \textbf{(c)} Decaying pattern in d$V$/d$I$ as a function of in-plane field applied \textit{across} junction A, \textit{along} direction of current flow.
    \textbf{(d)} Asymmetrically decaying pattern in d$V$/d$I$ as a function of in-plane field applied \textit{along} junction B, shifted by \SI{120}{degrees} in the sample plane with respect to junction A.
    }
\end{figure}

Now we turn to the magnetic field response of the device. It is instructive to first isolate the single-junction response from the two-dimensional bias maps shown in Figs.~\ref{fig:1}(c,d). 

By choosing an appropriate combination of bias currents along the arm-like feature associated with junction A, the current distribution in the device can be adjusted such that junction A remains dissipationless while junctions B and C are driven into a resistive state.
This procedure is illustrated by the dark green line in Fig.~\ref{fig:1}(d) and explained in more detail in the Supplemental Material (cf. Fig.~S3).
The experimentally measured differential resistance thus effectively only probes a single Josephson junction, i.e., the signal is dominated by the junction under investigation (A), while the other two junctions, B and C, to first approximation only contribute as passive resistive elements.
In Fig.~\ref{fig:2}, the bias current $I_\mathrm{A}$ denotes the net current through junction A, defined relative to the $I_\mathrm{B}=-I_\mathrm{C}$ line. Moving perpendicular to this line in bias space leads to a net current through junction A (see also Supplemental Material for details).
This approach provides a controlled way to isolate the magnetic-field response of an individual junction within the multiterminal geometry and serves as a basis for developing an intuitive understanding of the full three-terminal device's magnetic response, as discussed in Sec.~\ref{sec:threetfield}.

The magnetic field experienced by one junction is expressed in terms of three orthogonal components $B_\perp$, $B_\parallel$ and $B_{\rm I}$, where $B_\perp$ is the out-of-plane field, $B_\parallel$ is the in-plane field perpendicular to the current direction and $B_I$ is the in-plane field parallel to the current direction. 
While the out-of-plane field component B$_\perp$ is common for all junctions, the two in-plane field directions differ for each junction, cf. the insets in Fig.~\ref{fig:2}.
The procedure to align the device coordinate system with our magnet is outlined in the Supplemental Material.
Due to the geometric symmetry of the device, the three junctions are expected to exhibit qualitatively equivalent behavior; measurements performed on junction A are in the following treated as a representative example.

Fig.~\ref{fig:2}(a) shows the transport across junction A under an applied out-of-plane field.
This is a well-studied case, in which the magnetic vector potential induces a phase modulation along the junction, resulting in a Fraunhofer-like modulation of the critical current.
The particularly regular Fraunhofer pattern testifies to a homogeneous current distribution along the width of the junction, which confirms the validity of the side-contacted fabrication approach.
We find a periodicity of $\Delta B$ = \SI{5.5}{mT}, which is in good agreement with the geometrical proportions of junction A when considering flux-focusing effects \cite{Ghatak2018}.

Fig.~\ref{fig:2}(b) shows the junction's response to an in-plane field $B_{\parallel,\mathrm{A}}$ applied along the junction.
As the applied magnetic flux threads the area enclosed between the top and bottom surface of the TI nanowire, a nano-SQUID-like behavior is expected \cite{Nikodem2025}.
A periodicity of $\Delta B \approx \SI{380}{mT}$ is observed. For junction A, AFM and SEM measurements give $t \approx \SI{46}{nm}$ and $L \approx \SI{140}{nm}$. Assuming that the flux-enclosing area is reduced by the extent $z_0$ of the surface-state wave function into the bulk, $\Phi_0 = \Delta B\,(t - 2z_0)(L - 2z_0)$, the observed periodicity corresponds to $z_0 \approx \SI{3}{nm}$, comparable to previous results on this material system~\cite{Nikodem2025}.  
This pattern is a consequence of the bulk-insulating nature of our TI, resulting in the supercurrent being carried predominantly by the surface states.
From the shape of SQUID lobes, the $I_c$-asymmetry of the top and bottom surface can be inferred.
We find $I_{\rm c,\mathrm{1}} = \SI{225}{nA}$ and $I_{\rm c,\mathrm{2}} = \SI{75}{nA}$. 

Whereas in previous work the response to an applied gate voltage could be used to assign the critical currents to top and bottom surface, respectively \cite{Nikodem2025}, the more complicated gating behavior here (see Sec.~\ref{sec:gating}) prevents us from doing this with confidence.

Fig.~\ref{fig:2}(c) shows the junction's response to an applied in-plane field $B_I$ \textit{across} the direction of the junction, \textit{along} the direction of the current flow.
Above $\approx \SI{30}{mT}$, the maximum $I_{\rm c}$ is roughly linearly suppressed until the zero-resistance state ceases at $B \approx$ \SI{400}{mT}.
This decay is attributed to the orbital effect of the applied magnetic field. In a semiclassical picture, trajectories traversing the junction at an angle pick up an extra phase proportional to the field strength and their lateral displacement. Since this phase differs between trajectories, their contributions to the supercurrent partially cancel, resulting in a reduction of the critical current~\cite{Chen2018,Indolese2020}.

Fig.~\ref{fig:2}(d) shows the junction’s response to an in-plane field $B_{\parallel,\mathrm{B}}$ along junction B, which is rotated by \SI{120}{\degree} in-plane compared to the field direction in Fig.~\ref{fig:2}(b).
Junction A thus experiences a mixed field $B_\mathrm{mf,A}$, a superposition of $B_\parallel$ and $B_{\rm I}$. We find that the zero-resistance state vanishes even sooner than in Fig.~\ref{fig:2}(c).
This rapid decay is highly relevant for the device, since for any in-plane field $B_{\parallel,i}$, junctions j and k experience mixed fields $B_\mathrm{{mf,j}}$ and $B_\mathrm{{mf,k}}$, respectively. 
To check whether all the above magnetic-field responses can consistently be understood in a single-junction picture, we employ a semiclassical trajectory model based on Ref. \onlinecite{Chen2018}, where the critical current of a single junction at a given magnetic field is obtained by maximizing the sum over all possible trajectories weighted by their length and phase difference. For simplicity, we neglect the Zeeman term, whose impact was found to be much smaller than that of the phase modulation\cite{Chen2018}.
\begin{align}
\begin{split}
     I_c&\propto \max_{\phi_0}\int_{-W/2}^{W/2}dx_1\int_{-W/2}^{W/2}dx_2D\left [\sin\left(\phi_0+\phi_{\parallel}+\phi_{I}+\phi_{\perp}\right) \right.\\
    &+\left. \delta\sin\left(\phi_0-\phi_\parallel-\phi_I+\phi_\perp\right) \right]\\
    \end{split}
\end{align}
With
\begin{align*}
         D&=\frac{1 }{L^2+(x_1-x_2)^2},\\
    \phi_{\parallel}&=\pi \frac{B_{\parallel}tL}{\Phi_0},\\
    \phi_{\rm I}&=\pi\frac{B_{\rm I}(x_1-x_2)t}{\Phi_0},\\
    \phi_{\perp}&=\pi\frac{B_{\perp} L_{\mathrm{eff}}(x_1+x_2)}{\Phi_0}.
\end{align*}
Here $\Phi_0=\frac{h}{2e}$ is the superconducting flux quantum, $\delta$ is the asymmetry in critical current between the top and bottom junctions, $\phi_\parallel, \phi_{\rm I}, \phi_\perp$ describe the phase modulation imposed by the respective magnetic fields and $L_{\mathrm{eff}}$ is the effective junction length taking flux-focusing into account \cite{Ghatak2018}.  Note that the phase terms resulting from $B_\parallel$ and $B_I$ have different signs for the top and bottom junction, which can be understood from the direction of the screening currents. With this, we can calculate the critical current for the different magnetic field directions (see Supplemental Material for parameter values).  The results are superimposed as dashed lines on the experimental data in Fig.~\ref{fig:2}. Taking into account that this model neglects higher-harmonic contributions to the supercurrent and assumes a homogeneous geometry and supercurrent density, we find good semi-quantitative agreement between calculation and measurement, confirming that our experimental procedure is suitable to study the magnetic-field response of the individual junctions. Furthermore, it confirms that  considering only the topological surface state is sufficient to understand the supercurrent signatures in magnetic field, demonstrating again the bulk-insulating nature of our crystals.



\section{Three-terminal magnetic-field response}
\label{sec:threetfield}

\begin{figure}[t]
\includegraphics[width= 3.35in]{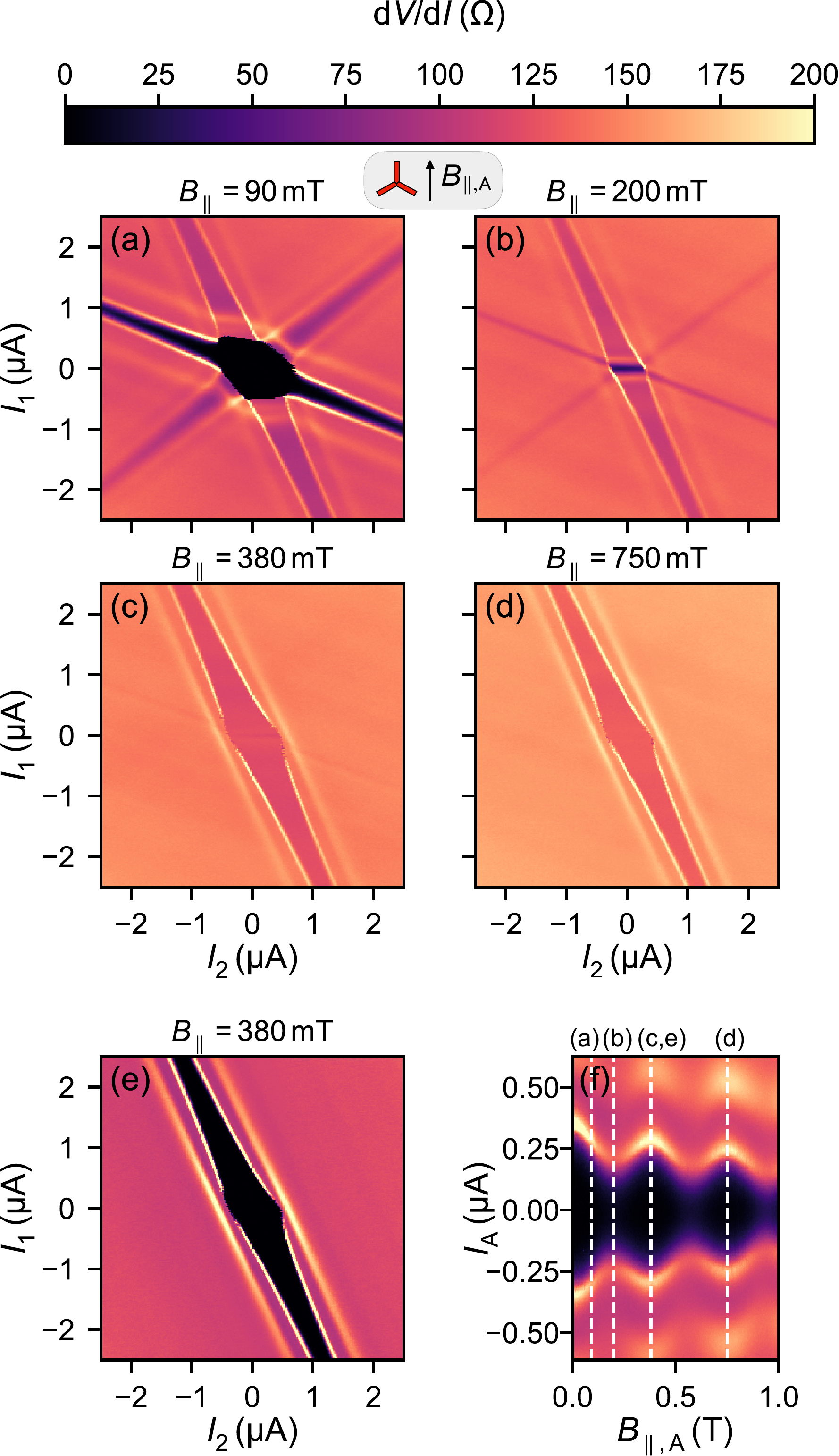}
\caption{
    \label{fig:inplane}
    Three-terminal response to $B_{\parallel,{\rm A}}$. Inset shows field working points in relation to SQUID pattern of Fig.~\ref{fig:2}(b). 
    Panels (a--d) show d$V_{10}$/d$I_{1}$, (e--f) show d$V_{20}$/d$I_{2}$.
    \textbf{(a)} $I_{\rm c}$ reduction slowly setting in, shape of CCC slightly affected.
    \textbf{(b)} Zero-resistance features of B and C nearly vanished, A at oscillation minimum.
    \textbf{(c)} 
    Junction A nearly fully recovered at maximum of first SQUID lobe.
    \textbf{(d)} Simultaneous SQUID oscillation on all junctions.
    Features of B and C are not visible because of the $B_{\rm I}$ contribution. 
    \textbf{(e)}
    d$V_{20}$/d$I_{2}$ at $B_\parallel$=380 mT, showing that junction A is still superconducting, but the CCC has vanished.
    \textbf{(f)}
    Nano-SQUID pattern of junction A with white dashed lines indicating the magnetic-field values at which the bias maps in panels (a–d) were recorded. 
    }
\end{figure}

\begin{figure}[t]
\includegraphics[width= 3.35in]{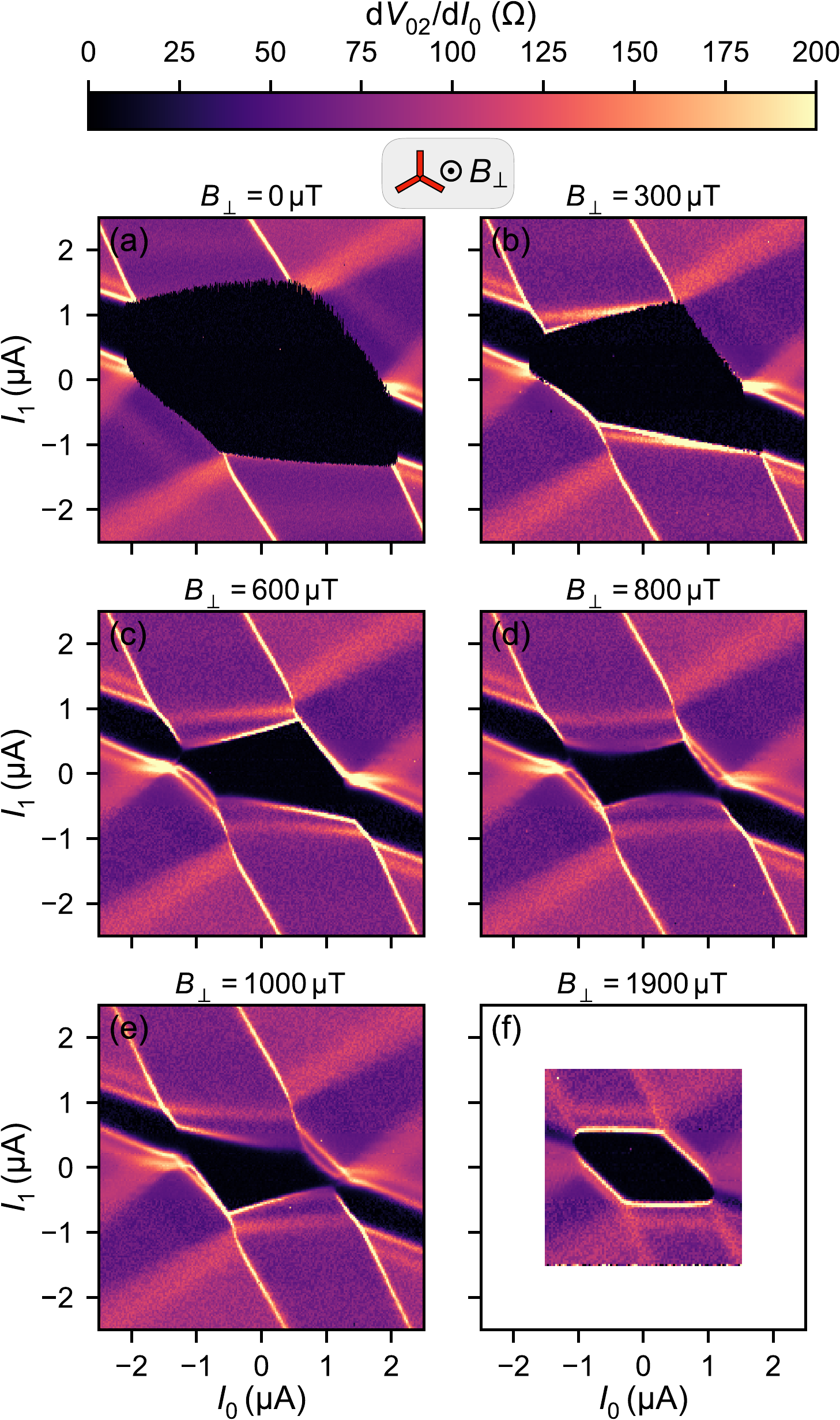}
\caption{
    \label{fig:oop}
    Three-terminal response of device 2 to small $B_{\perp}$, showing a cyclic deformation of the CCC.  While the individual critical currents of the junctions remain largely unaffected, an imposed phase condition between terminals leads to deformations of the CCC.
    \textbf{(a)} Zero-field bias map.
    \textbf{(b)} The deformation sets in on the left half of the contour, extends to the
    right half \textbf{(c)}, and is strongest at \SI{800}{\micro\tesla} \textbf{(d)}, where the
    contour becomes non-convex and diamond-shaped.
    It then relaxes, again first
    on the left half \textbf{(e)}.
    \textbf{(f)} At \SI{1900}{\micro\tesla}, the CCC is restored in a shape similar to, but smaller than, the zero-field case, resulting from an overall reduced $I_\mathrm{c}$.}
\end{figure}

After establishing the fundamental response of the individual junctions to an applied magnetic field, we next investigate whether the field simply modulates each junction independently or also affects the collective superconducting state of the device. 
We therefore study the full bias maps under applied fields $B_\parallel$ (cf. Fig.~\ref{fig:inplane}) and $B_\perp$ (cf. Fig.~\ref{fig:oop}), with particular emphasis on the deformation of characteristic features and changes in the shape of the CCC.

Figure \ref{fig:inplane} shows the evolution of the full bias map as a function of an increasing in-plane magnetic field $B_\parallel$ applied along junction A, $B_{\parallel,\mathrm{A}}$, characterized by measuring d$V_{10}$/d$I_{1}$.
Owing to the rotational symmetry of the device, equivalent behavior is obtained upon rotating the field to lie along junctions B or C.
Fig.~\ref{fig:inplane}(e) shows d$V_{20}$/d$I_{2}$ at $B_\parallel=\SI{380}{mT}$ for comparison, while Fig.~\ref{fig:inplane}(f) serves as a visual guide indicating the fields at which Figs.~\ref{fig:inplane}(a--d) were recorded within the SQUID pattern.

For small $B_\parallel$ (a), the bias map remains largely unchanged.
Only minor modifications of the CCC are observed, which are attributed to a small parasitic out-of-plane field component $B_\perp$ arising from sample misalignment.

At $B_{\parallel,\mathrm{A}} \approx \SI{200}{mT}$ (b), consistent with the behavior shown in Fig.~\ref{fig:2}(b,d), the critical currents of junctions B and C are strongly suppressed, while junction A is penetrated by a flux of $\frac{\Phi_0}{2}$ along the junction and thus remains superconducting close to a minimum of its SQUID-like oscillation. Consequently, the size of the CCC is strongly reduced. 
Increasing the field to $B_{\parallel,\mathrm{A}} \approx \SI{380}{mT}$, shown in Fig.~\ref{fig:inplane}(c), restores $I_{\rm c,\mathrm{A}}$, whereas junctions B and C are now fully resistive. Accordingly, the CCC has vanished. Dissipationless transport from terminal 1 to terminal 2 is no longer possible, but junction A can still carry a supercurrent as is visible from the anti-diagonal feature in Fig.~\ref{fig:inplane}(c,e). 

When the field is increased further to $B_{\parallel,\mathrm{A}}$ $\approx$ \SI{750}{mT} (Fig.~\ref{fig:inplane}(d)), junctions B and C are penetrated by a $B_\parallel$ corresponding to $\Phi_0$, but consistent with the data shown in Fig.~\ref{fig:2}(d), the $B_I$ component of the magnetic field prevents a recovery of the supercurrent of these junctions. In contrast, junction A sees a $B_\parallel$ corresponding to  $2\Phi_0$ and no $B_I$ and therefore remains superconducting.

The observed bias map is readily understood in terms of the single-junction response and the vanishing CCC in Fig.~\ref{fig:inplane} reflects their reduced critical currents.

This behavior changes drastically when we consider the response of the trijunction to a small out-of-plane magnetic field $B_\perp$.
In the following, we discuss a second device that is lithographically equivalent to device 1 but with larger junction widths of approximately \SIrange{1}{1.5}{\micro\meter}, resulting in higher critical currents and therefore improved visibility of the transport features.
For device 2, terminal 2 is grounded and bias currents $I_0$ and $I_1$ are applied.
A full characterization of device 2 as well as the full out-of-plane magnetic-field response of device 1 is provided in the Supplemental Material.

Figure \ref{fig:oop}(a) shows the zero-field bias map of device 2.
The strongly enlarged CCC compared with device 1 is a consequence of the substantially larger critical currents of this device.
The characteristic Fraunhofer-oscillation period of a single junction on this device is $\approx$ \SI{3}{mT} (cf. Supplemental Material Fig.~S6(b)), which is consistent with the geometric junction dimensions.
Concerning the full trijunction, applying an out-of-plane magnetic field produces a modulation of the CCC that we term the three-terminal analogue of a Fraunhofer interference pattern.
This is shown in the bias maps of Fig.~\ref{fig:oop}(b--f), where increasing out-of-plane fields $B_\perp$ are applied.

The out-of-plane field progressively distorts the CCC from its zero-field shape into a strongly non-convex, diamond-like contour, most pronounced around \SI{800}{\micro\tesla}.
This field approximately corresponds to the penetration of a single magnetic flux quantum through the effective area of the entire trijunction, including flux-focusing effects.
Remarkably, upon further increasing the magnetic field, the deformation gradually reverses and evolves back in an approximately inversion-symmetric manner with respect to the field values below the maximum deformation. 
Clearly, the shrinking and deformation of the CCC cannot be understood as a consequence of the magnetic field affecting the individual junctions.
At small fields, only marginal suppression of the critical currents is expected, as can be seen from the width of the corresponding arms and is also understood from Fig.~\ref{fig:2}(a).


To deepen the understanding of the  magnetic-field evolution of the bias maps, we model the device as a triangular network of three pairwise Josephson junctions within an extended RCSJ framework, based on the code published by Gupta et al. \cite{Gupta2023}.
We account for the application of a perpendicular magnetic field in two ways.
First, the critical current of each individual junction is reduced by a Fraunhofer-like field dependence.
Second, the magnetic flux $\Phi_\mathrm{tri}$ threading the total effective area of the device imposes the condition $\phi_{\rm A}+\phi_{\rm B}+ \phi_{\rm C}=2\pi\frac{\Phi_\mathrm{tri}}{\Phi_0}$ on the phase differences $\phi_i$ of the respective junctions, similar to the situation of a conventional SQUID. We therefore explicitly neglect the phase buildup inside the individual junctions.
Strictly speaking, these assumptions are valid only if the total loop area is large compared to the individual junction areas, so we limit ourselves to below one flux quantum in the loop
(see Supplemental Material Sec. VII for details including simulation parameters).

\begin{figure}
    \centering
    \includegraphics{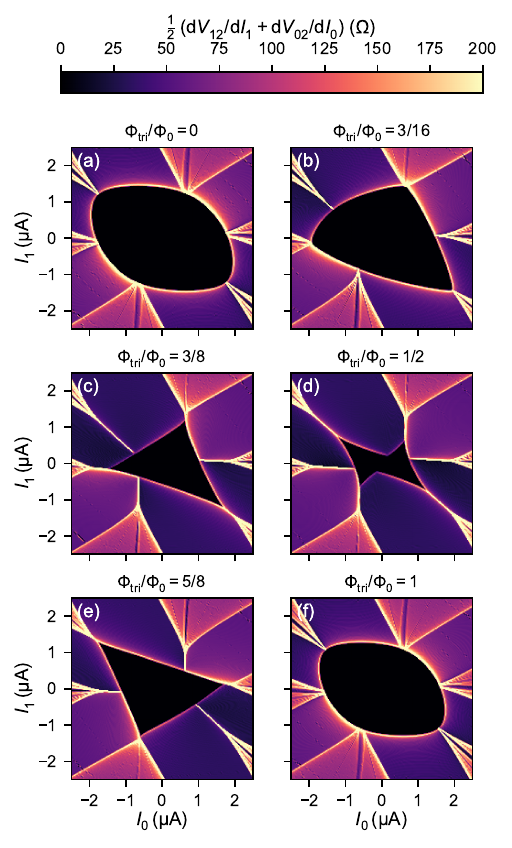}
    \caption{Simulated out-of-plane magnetic-field response for device 2 using extended RCSJ model including phase frustration.
    The full set of parameters can be found in the Supplemental Material.
    CCC deformation increases continuously with increasing magnetic flux and reaches its maximum at $\Phi_{\mathrm{tri}}=\Phi_0/2$, where the CCC approaches a diamond-like shape.
    Upon increasing the flux further, the non-convex deformation gradually decreases again.
    \textbf{(a)} Zero-field bias map.
    \textbf{(b--f)} Increasing phase frustration up to maximum at $\Phi_\mathrm{tri}/\Phi_0 = 1/2$, decreasing thereafter.
    }
    \label{fig:sim}
\end{figure}

The results shown in Fig.~\ref{fig:sim} are then obtained by solving the coupled RCSJ equations numerically for each pair of applied bias currents $(I_0,I_1)$. To better identify the exact shape of the CCC, we plot the average resistance of both terminals to ground, which allows for clear recognition of all relevant features.
Fig.~\ref{fig:sim}(a) shows the zero-field bias map, and Figs.~\ref{fig:sim}(b--f) show the deformation of the CCC closely resembling the evolution of Fig.~\ref{fig:oop}.
Importantly, this model does not contain explicit non-pairwise Josephson couplings in the form of, e.g., a central scattering region \cite{Pankratova2020}, yet is able to reproduce the non-convex deformation of the CCC.

The maximum deformation occurs within the model at a flux penetration of $\Phi_\mathrm{tri} = \Phi_0/2$, while the device shows maximum deformation at $\Phi_\mathrm{tri} = \Phi_0$ including flux-focusing effects, which might be explained by the device not forming a single well-defined loop but rather an ensemble of loops of different sizes. A better quantitative understanding of this problem is beyond the scope of this work. 
We conclude that device~2 is well described by three pairwise coupled Josephson junctions within the RCSJ model; this neither requires nor excludes minor non-pairwise three-terminal contributions.  
The out-of-plane field dependence recorded on device~1 (see Supplemental Material) also shows a flux-dependent modulation, which our model largely reproduces. However, the non-convex deformation that the model predicts for device~1 as well is not observed experimentally. 


 \section{Electrical Gating Behavior}
\label{sec:gating}
Finally, we investigate the response of device 1 to electrostatic gating. This enables us to probe whether a change in chemical potential modifies the superconducting features uniformly throughout the trijunction.   
Using the degenerately doped Si substrate as a global back gate allowed us to reliably apply gate voltages of up to $\pm$\SI{40}{V}. 

\begin{figure}[t]
\includegraphics[width= 3.35in]{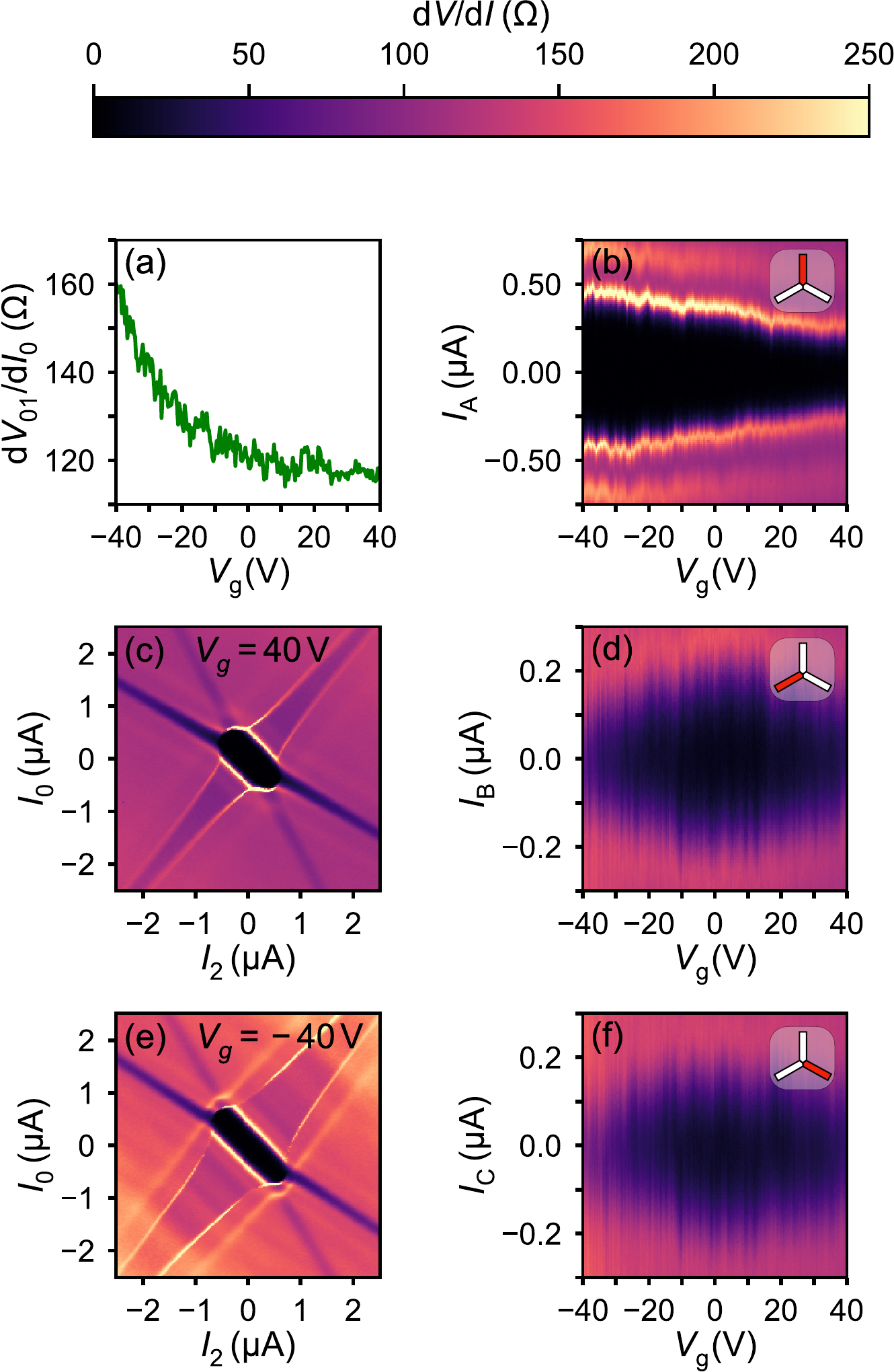}
\caption{
    \label{fig:gating}
    Electrical gating behavior of the trijunction.
    Gate voltage applied via a global back gate. 
    Insets indicate the measured junction in red.
    \textbf{(a)} Gate-dependent normal state  resistance $\mathrm{d}V_{01}/\mathrm{d}I_0$.
    \textbf{(b,d,f)} $I_{\rm c}$ versus $V_{\rm g}$ behavior of junction A (b), junction B (d) and junction C (f) measured along line cuts comparable to the ones used in Fig.~\ref{fig:2}.
    \textbf{(c,e)} Full two-dimensional bias characterization at (c) $V_{\rm g}$ = \SI{40}{V} and (e) $V_{\rm g}$ = \SI{-40}{V}, d$V_{01}$/d$I_{0}$ measured in a different grounding configuration (terminal 1 grounded, cf. Fig.~S1(c,d)).
    }
\end{figure}

In Fig.~\ref{fig:gating}(a), the differential resistance $\mathrm{d}V_{01}/\mathrm{d}I_0$ measured in the non-superconducting region E is plotted as a function of gate voltage $V_{\rm g}$. 
We observe a monotonic increase in resistance with more negative gate voltages, indicating that the TI is in an n-type regime, where a more negative gate voltage reduces the chemical potential towards the Dirac point.
Within the gate range available to us, no maximum in resistance was reached, so we can conclude that we did not cross the Dirac point. This could be a consequence of our flake being thicker than the ones in previous experiments, where the Dirac point was observed \cite{Nikodem2025,Nikodem2025diode}.

Figs.~\ref{fig:gating}(b), \ref{fig:gating}(d), and \ref{fig:gating}(f) show the evolution of $I_{\rm c}$ over the applied $V_{\rm g}$ for junctions A, B, and C, respectively, measured in the same way as described in Sec.~\ref{sec:singlefield}. 
Junction A shows an increase of $I_{\rm c}$ for negative gate voltages, whereas junctions B and C show the expected decrease in critical current for negative $V_{\rm g}$ and little tunability for positive gate voltages.

Indeed, the full two-dimensional current-bias maps at $V_{\rm g}=\SI{40}{V}$ (Fig.~\ref{fig:gating}(c)) and $V_{\rm g}=\SI{-40}{V}$ (Fig.~\ref{fig:gating}(e)) confirm this observation. The arm corresponding to junction A indeed broadens for more negative gate voltages, while the arms of junctions B and C show a decrease in width. 

\section{discussion}
\label{sec:discussion}
The device realized here effectively comprises two trijunctions, one at the top and one at the bottom surface of the TI.
Our magnetic-field-dependent measurements of individual junctions combined with semiclassical modeling confirm this picture, highlighting the negligible role of possible bulk states for the supercurrent.
The in-plane field dependence of the two-dimensional bias characterization can be qualitatively understood by considering three individual nano-SQUID-like junctions. Intuitively, this follows from the locality of the phase buildup enforced by in-plane magnetic fields. 
However, the deformation of the CCC under an out-of-plane field cannot be described this way. The reason lies in the fluxoid quantization enforced over the closed loop formed by the trijunction, which sets a constraint on the allowed phase combinations, affecting all three junctions simultaneously.
By modeling our device as a network of three pairwise coupled RCSJ junctions, we were able to qualitatively reproduce the out-of-plane dependence of the two-dimensional bias maps observed on device~2.
The response of device~1 to an applied back-gate voltage, on the other hand, could not be captured by these models. In the following we speculate about its possible origin.
It was noted early on that there is a qualitative difference between a trijunction consisting of three pairwise coupled junctions and a fully coherent one \cite{Pankratova2020}.
Tuning the gate could therefore modify the ABS spectrum of the trijunction in a non-trivial way, resulting in the observed unexpected gate dependence.
Interestingly, a similar effect was already observed in Ref. \onlinecite{Graziano2022} (see Fig.~3 and Fig.~S4 there), where the authors had to increase the critical currents for ungated segments of their semiconductor junction upon depleting others to successfully model their experimental results.

Alternatively, the observed back-gating response may arise from spatial variations of the chemical potential.
Since a topological insulator is a Dirac material, such fluctuations can place different regions of the trijunction on opposite sides of the Dirac point \cite{Borgwardt2016, Knispel2017}.
However, in all our previous experiments, the nanodevice fabrication led to clear n-type behavior \cite{Ghatak2018,Schluck2024,Nikodem2025,Nikodem2025diode}.
Even if segments of the junction were in the p-type regime, the transparency to the electrodes would typically be strongly reduced, resulting in a saturation of the critical current rather than an increase with gate \cite{Schmidt2018, Nikodem2025}. Furthermore, for the semiconducting platform of Ref. \onlinecite{Graziano2022}, this explanation does not apply.
We therefore conclude that the observed behavior is most likely associated with the trijunction geometry itself, potentially involving a genuine multiterminal contribution beyond purely pairwise coupling.

In conclusion, we have successfully demonstrated the suitability of the side-contacted TI platform for fabricating high-quality trijunctions, whose supercurrent is carried by the topological surface state.
We find that the response to an in-plane field can qualitatively be described in a semiclassical single-junction picture, whereas fluxoid quantization couples all three junctions when exposed to an out-of-plane field.
However, the gate response remains unexplained by these considerations and could hint at a true multiterminal effect, while other explanations cannot be fully excluded from the data alone.
It is therefore conceivable that at least some ABSs in the device couple to all three superconducting phase parameters, which is a prerequisite for the discrete vortex underlying the Fu--Kane trijunction scenario in such a device geometry.
Since both surfaces in this device form a trijunction, a phase vortex would generically bind a zero mode on each surface; how the coupling between top and bottom surface affects the ABS spectrum, and whether it hybridizes these modes, is therefore a key question for future studies.

\begin{acknowledgments}
This work has received funding from the Deutsche Forschungsgemeinschaft (DFG, German Research Foundation) under CRC 1238 - 277146847 (Subprojects A04 and B01) and from DFG under Germany's Excellence Strategy - Cluster of Excellence Matter and Light for Quantum Computing (ML4Q) EXC 2004/2 - 390534769.
The authors also acknowledge support from the DFG Major Instrumentation Programme under project No. 544410649.
The authors used Claude (Anthropic) to review the manuscript for internal consistency and clarity and to suggest revisions to the wording. All suggestions were critically assessed and, where adopted, revised and verified by the authors, who take full responsibility for the content of this article.
\end{acknowledgments}
\section*{DATA AVAILABILITY}
The data and simulation codes that support the findings of this article are openly
available under the Zenodo repository with the identifier 10.5281/zenodo.23562455.



\bibliography{bib}


\end{document}


\preprint{APS/123-QED}

\title{Supplementary information to Magnetotransport experiments on side-contacted topological insulator three-terminal Josephson junctions}

\author{Christoph Wassmer}
\author{Ella Nikodem}
\author{Oliver Breunig}
\author{Yongjian Wang}
\author{Yoichi Ando}
\author{Jakob Schluck}
\email{schluck@ph2.uni-koeln.de}
\affiliation{
 Physics Institute II, University of Cologne, Zülpicher Straße 77, 50937 Köln, Germany
}
\date{\today}


\maketitle


\section{Alternative grounding configurations}

\begin{figure}[t]
\includegraphics[width= 3.35in]{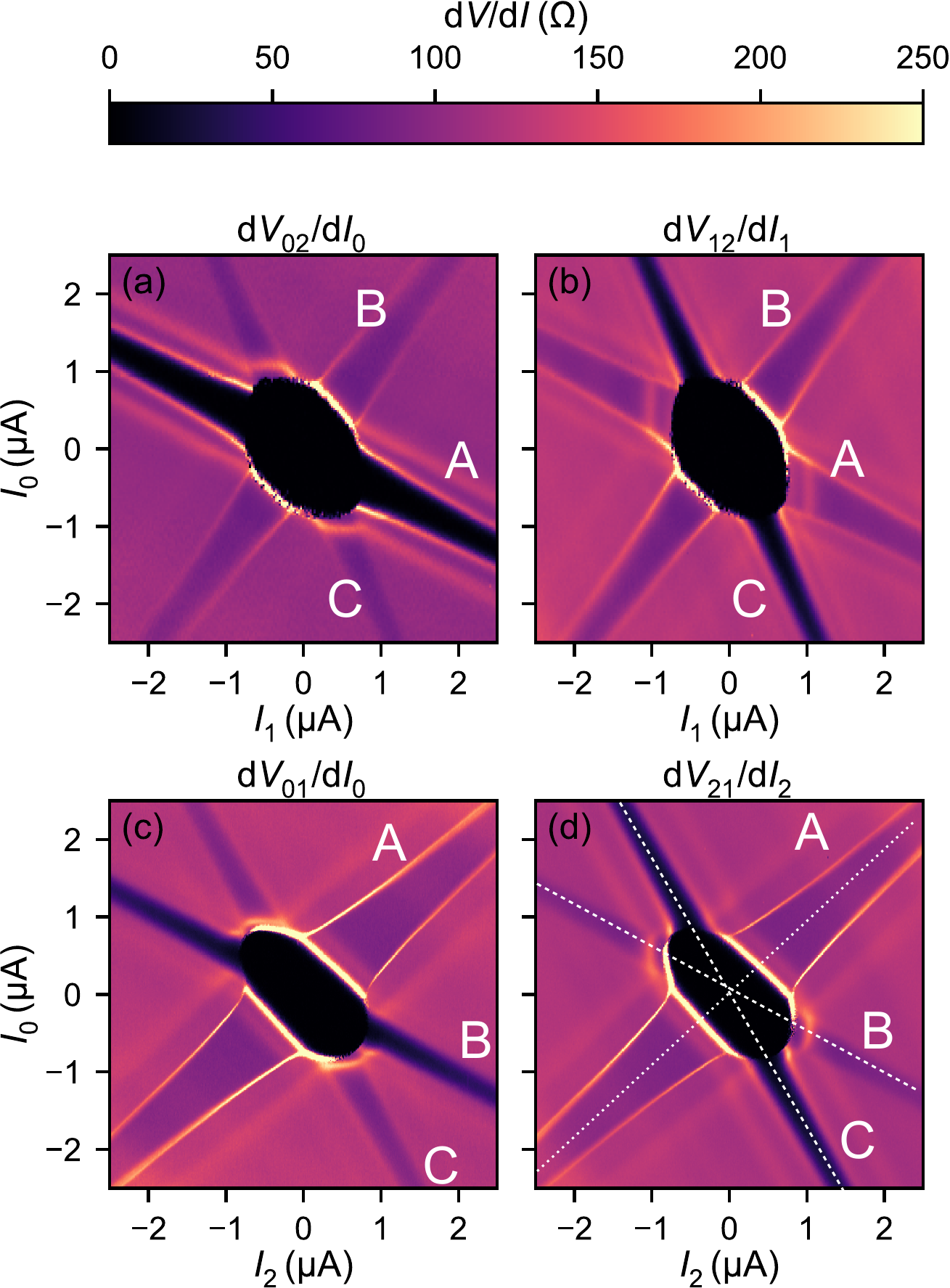} 
\caption{
    \label{fig:1} 
    Bias maps of device 1 measured at zero magnetic field using different grounding configurations. Features are labeled according to the corresponding junction.
    \textbf{(a,b)} Terminal 2 is grounded.
    \textbf{(c,d)} Terminal 1 is grounded.
    }
\end{figure}

Fig. \ref{fig:1} shows the zero-field bias maps corresponding to Figs.~1(c,d) of the main text, measured using alternative grounding configurations.
In Figs. \ref{fig:1}(a,b) terminal 2 is grounded, while terminals 0 and 1 are biased, whereas in (c,d) terminal 1 is grounded, while terminals 0 and 2 are biased.
The first index of d$V_{ij}$/d$I_{i}$ denotes the biased terminal and the second index the grounded terminal.
Changing the grounded terminal permutes the positions and changes the slope of the features associated with the individual junctions.
The feature belonging to the junction spanned by the biased terminals (junction B in Figs. \ref{fig:1}(a,b) and junction A in Figs. \ref{fig:1}(c,d)) will appear double the width of the other features.
A single grounding configuration already provides a complete physical description of the device. Changing the grounded terminal therefore does not affect the underlying physics in magnetic fields.

The width of the star-shaped features does not directly reflect the intrinsic critical current of the respective junction because Joule heating in the two resistive junctions substantially increases the electron temperature in the shared weak link area, thereby reducing the observed switching current.

\section{Magnetic field alignment}

\begin{figure}[t]
\includegraphics[width= 3.35in]{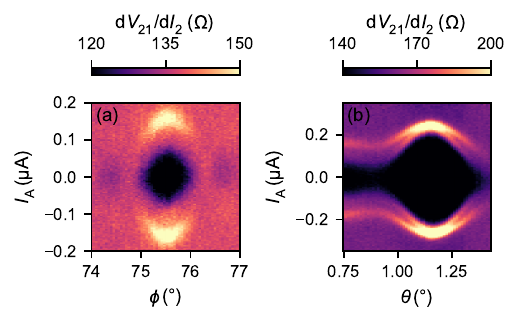}
\caption{
    \label{fig:magnetalignment} 
    Magnetic field alignment process recorded on device 1, junction A.
    \textbf{(a)} In-plane (in direction across junction A) / out-of-plane alignment
    \textbf{(b)} In-plane (in direction of junction A) / out-of-plane alignment
    }
\end{figure}

To ensure that the magnetic field directions are accurately defined, the coordinate system of the superconducting magnet was aligned with that of the sample.
The measurements used for this alignment are shown in Fig. \ref{fig:magnetalignment}.
Junction A is used as a reference and is measured along the green cut shown in the main text.
Here we use the standard convention, in which $\theta$ is the polar angle measured from the $z$ axis and $\phi$ the azimuthal angle measured from the $x$ axis within the $xy$ plane.

First, a $\phi$ sweep (Fig. \ref{fig:magnetalignment}(a)) is performed at a constant field magnitude of $B_\rho$ = \SI{300}{mT}, varying the magnetic field orientation between out-of-plane and in-plane (across junction A) components.
Any contribution of out-of-plane field drastically reduces the critical current, so only when the field is aligned close to the in-plane direction at $\phi \approx \SI{75.6}{\degree}$ the critical current is determined predominantly by the in-plane field component in current direction.

Fig. \ref{fig:magnetalignment}(b) shows the alignment of the $\theta$ angle between the out-of-plane and in-plane (along junction A) components.
At a high field magnitude of $B_r $= \SI{1}{T}, even small deviations from the in-plane orientation result in a strong reduction of the critical current.

The influence of slight misalignment between the two in-plane field directions was investigated and found to be negligible.

\section{Single junction measurement path}

\begin{figure}[t]
\includegraphics[width= 3.35in]{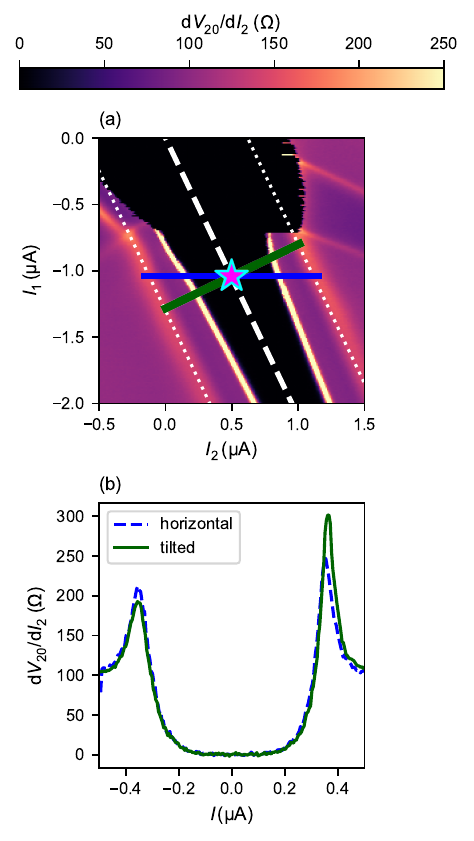}
\caption{
    \label{fig:path} 
    Measurement method for isolated junction data
    \textbf{(a)} Zoom-in of main text Fig.~1(d) bias map, tilted measurement path in dark green, alternative horizontal measurement path in dark blue
    \textbf{(b)}
    Same d$V$/d$I$-cut measured along blue / green path, green curve stretched by geometric factor 1.10.
    }
\end{figure}

The single-junction measurements shown in Fig. 2 of the main text are obtained by isolating one junction in the superconducting state through measurements along a selected path in bias space.
Fig. \ref{fig:path}(a) shows the measurement path used in dark green.
It is chosen such that it intersects the zero bias current configuration of junction A (white dashed line) at a \SI{90}{\degree} angle, minimizing asymmetric contributions from Joule heating in the resistive junctions.
The measurement path can be mathematically expressed by a line of slope
$$
m_\mathrm{green} = -\frac{1}{m_\mathrm{wd}}
$$
where $m_\mathrm{wd}$ is the slope of the white dashed line given by
$$
m_\mathrm{{wd}} = -\left(\frac{R_B}{R_C}+1\right)
$$
where $R_B,R_C$ are the normal-state resistances of junctions $B$ and $C$.

Measurements performed along an alternative horizontal path (blue) yield the same behavior after rescaling the current axis by the geometrical factor determined by the angle between the two paths (Fig. \ref{fig:path}(b)).

The magenta star marks the intersection between the zero-current configuration of junction A and the two measurement paths. This bias configuration corresponds to the measurement point used for Fig. 6(b).
The $I_\mathrm{c}$ gating behavior shown in Figs. 6(d,f) of the main text is obtained using the same approach, with the measurements performed along the respective features.



\section{Details of semiclassical numerical calculation}

Equation 1 of the main text is a natural generalization of the model discussed in Ref. \onlinecite{Chen2018} for a side-contacted TI, considering contributions from both surfaces to the supercurrent, while ignoring the bulk. While the out-of-plane component affects both surfaces equally, the phases due to in-plane fields build up asymmetrically. To keep the model simple, we consider identical geometries for top and bottom junction as well as a homogeneous supercurrent contribution and only first harmonics terms.
The parameters used for the calculation are as follows:
\begin{align*}
    L&=\SI{134}{nm}\\
    L_\mathrm{eff}&=\SI{500}{\nano\meter}\\
    W&=\SI{0.8}{\micro\meter}\\
    t&=\SI{40}{nm} \\
    \delta&=3    
\end{align*}
All of these values are in good agreement with the device geometry when one takes the extent of the surface state and flux-focusing into account.



\section{Device 1 Out-of-plane magnetic field bias maps}

\begin{figure}[t]
\includegraphics[width= 3.35in]{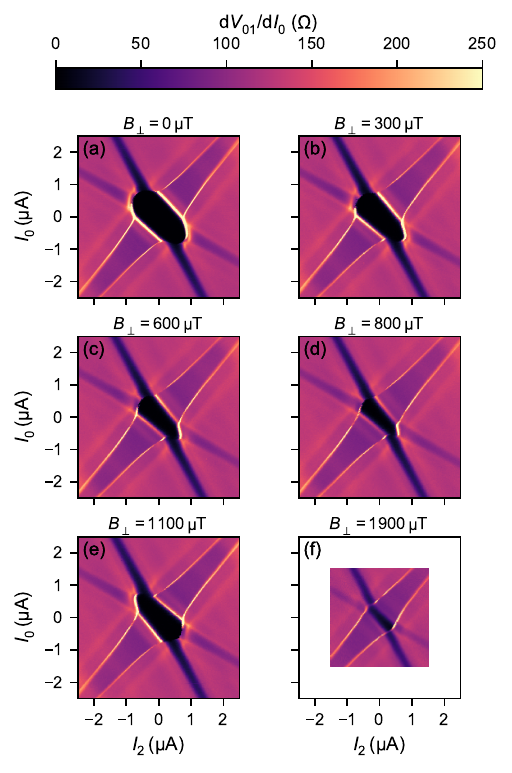}
\caption{
    \label{fig:device1oop} 
    Three terminal response of device 1 to small $B_\perp$.
    Cyclic deformation of the CCC can be observed, but much less pronounced than for device 2.
    The non-convex regime clearly observed in device 2 is not found here. 
    }
\end{figure}

\begin{figure}[t]
\includegraphics[width= 3.35in]{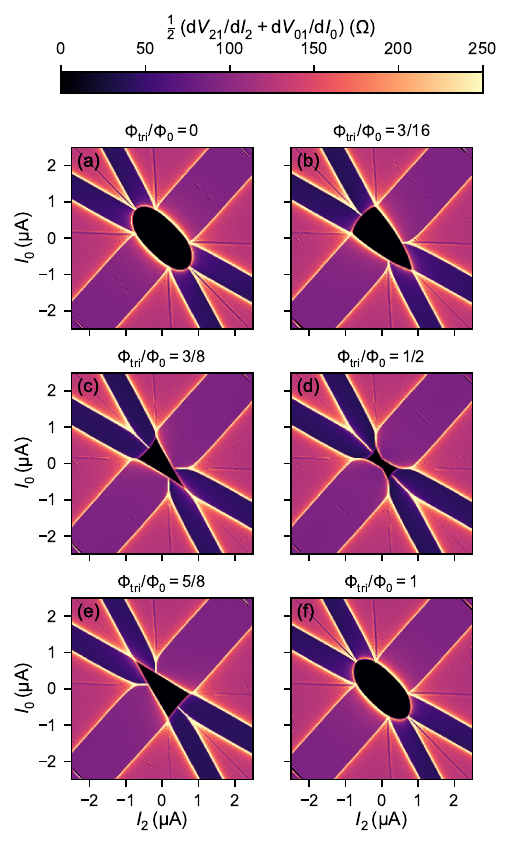}
\caption{RCSJ simulation of the response of device 1 to an out-of-plane magnetic field for flux through the whole trijunction area between 0 and $\Phi_0$. Simulation parameters can be found in Table~\ref{tab:D2_sim_parameters}.
    \label{fig:SimD1} 
    }
\end{figure}

Fig. \ref{fig:device1oop} shows the three terminal response to small out-of-plane fields $B_\perp$ recorded on device 1.
While CCC deformations are observed, their magnitude is less pronounced than in device 2 and does not reach the non-convex regime.
This is in contrast to our numerical simulations using the same RCSJ model as discussed in the main text and below. In Fig.~\ref{fig:SimD1}, we show the evolution of the full current bias map over the course of one flux quantum penetrating the trijunction area. Similar to the case of device 2 discussed in the main text, we find a cyclic deformation with a non-convex regime near half a flux quantum. 
Other than that, the general shape of the CCC deformation predicted for this combination of critical currents, from an ellipse to a triangle that changes direction upon increasing flux and reverts to an elliptical shape, is also found in the experiment.



\section{Device 2 main text measurements}

\begin{figure*}[t]
\includegraphics[width= 6.85in]{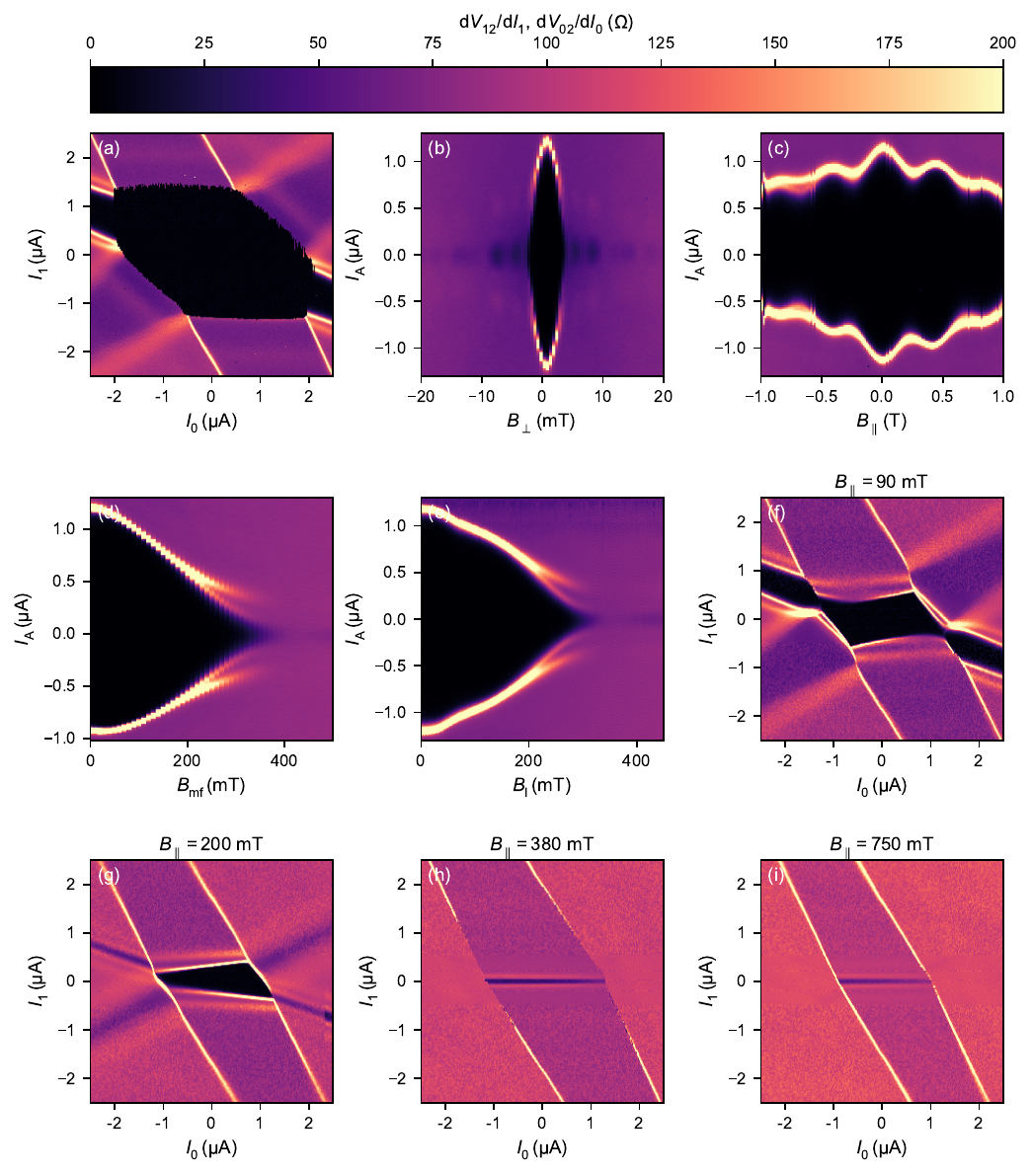}
\caption{
    \label{fig:device2} 
    Device~2 measurements corresponding to the experiments presented in Figs.~2 and 3 of the main text.
    \textbf{(a)} Zero-field bias map.
    \textbf{(b)--(e)} Single-junction characterization, including Fraunhofer interference, nano-SQUID oscillations, and the evolution of the critical current under magnetic field.
    \textbf{(f)--(i)} Bias maps measured for increasing in-plane magnetic field applied parallel to junction A, $B_{\parallel,\mathrm{A}}$.
}
\end{figure*}

Figure \ref{fig:device2} presents the measurements corresponding to Figs. 2 and 3 of the main text, performed on device 2.

Fig. \ref{fig:device2}(a) shows the zero field bias map.
The critical currents are larger in this device due to the larger junction widths of approximately \SIrange{1}{1.5}{\micro\meter}.

Fig. \ref{fig:device2}(b) shows the Fraunhofer interference pattern obtained by applying an out-of-plane magnetic field. 
The regularity of the pattern is again consistent with a homogeneous current distribution across the junction.
The periodicity is $\Delta B \approx\, $\SI{3}{mT}, which is slightly smaller than for device 1 due to the larger junction widths.

Fig. \ref{fig:device2}(c) shows the nano-SQUID recorded on junction A.
The inferred period is $\Delta B \approx \SI{415}{mT}$.
This is due to a thinner flake.
The asymmetry between the top and bottom TI surfaces is more pronounced than in device 1.

Fig. \ref{fig:device2}(d) shows the evolution of the critical current recorded on junction A under the application of an in-plane field along junction B $B_\mathrm{mf} = B_{\parallel,\mathrm{B}}$.

Figs. \ref{fig:device2}(f-i) show the evolution of the bias maps under an in-plane magnetic field parallel to junction A.
Since Fig. \ref{fig:device2}(f) closely resembles the CCC-pattern observed in Fig. 4(d), the amount of uncompensated out-of-plane component can be estimated to be approximately \SI{800}{\micro\tesla}. However, from the SQUID pattern shown in Fig.~\ref{fig:device2}(c), we know that this cannot be the result of a misalignment angle, as the resulting out-of-plane field would quickly suppress the critical currents of the individual junctions, which is not observed in the SQUID. We therefore speculate that there is a small remanent field of the superconducting magnet for this sweep that is sufficient to distort the CCC, but does not fully suppress the critical current of the individual junctions. 
Interestingly, in Figs. \ref{fig:device2}(h,i) the CCC does not completely vanish, even though the star-like features of reduced resistance indicating possible zero-resistance transport with junctions B and C are barely visible. 
In general, this device exhibits larger hysteresis in the critical current, which we attribute to increased Joule heating coming from the wider junctions. 

\section{RCSJ Simulation details}

We model our trijunction as a network of three pairwise Josephson junctions within an extended RCSJ framework starting from 
\begin{align}
I_0 &= -(I_1+I_2)\\
I_i &= I_{R,i} + I_{C,i} + I_{J,i}.
\end{align}
Assuming a sinusoidal CPR, the Josephson contributions $I_{J,i}$ to the junction currents can be described by 
\begin{align}   
I_{\mathrm{J},1} &= I_{\mathrm{c},12}\sin\phi_{12} + I_{\mathrm{c},10}\sin\phi_{10},\\
I_{\mathrm{J},2} &= - I_{\mathrm{c},12}\sin\phi_{12}+I_{\mathrm{c},20}\sin\phi_{20}.
\end{align}
where $\phi_{ij}$ is the phase difference between two terminals.
Applying a perpendicular magnetic  field has a flux threading the total effective area of the device,
\begin{equation}
    \Phi_{\mathrm{tri}} = B A_{\mathrm{tri}}^{\mathrm{eff}},
\end{equation}
and thus imposes a finite phase winding around the trijunction.

\begin{table*}[h]
    \centering
    \caption{Parameters used for the extended RCSJ simulation.}
    \label{tab:D2_sim_parameters}
    \begin{ruledtabular}
    \begin{tabular}{cccccccccc}
        Device &
        $I_{c,A}$ &
        $I_{c,B}$ &
        $I_{c,C}$ &
        $R_{A}$ &
        $R_{B}$ &
        $R_{C}$ &
        $C_0$ &
        $A_{ij}^{\mathrm{eff}}$ &
        $A_{\mathrm{tri}}^{\mathrm{eff}}$
        \\
        \hline
        1 &
        $\SI{500}{nA}$ &
        $\SI{325}{\nano A}$ &
        $\SI{375}{nA}$ &
        $\SI{145}{\ohm}$ &
        $\SI{200}{\ohm}$ &
        $\SI{170}{\ohm}$ &
        $\SI{5}{fF}$ &
        $\approx\SI{0.4}{\micro m^2}$ &
        $\approx\SI{1.5}{\micro m^2}$
        \\
        \hline
        2 &
        $\SI{650}{nA}$ &
        $\SI{1300}{\nano A}$ &
        $\SI{800}{nA}$ &
        $\SI{120}{\ohm}$ &
        $\SI{80}{\ohm}$ &
        $\SI{140}{\ohm}$ &
        $\SI{10}{fF}$ &
        $\approx\SI{0.7}{\micro m^2}$ &
        $\approx\SI{2.5}{\micro m^2}$
    \end{tabular}
    \end{ruledtabular}
\end{table*}

This phase winding frustrates the three pairwise Josephson couplings, since their gauge-invariant phase differences cannot be chosen independently anymore.
By choosing terminal 0 as the phase reference, $\phi_0=0$, the flux-induced phase shift is assigned to the junction connecting terminals
1 and 2.
The gauge-invariant phase differences are then
\begin{align}
    \phi_{12} &= \phi_1-\phi_2+2\pi\frac{\Phi_\mathrm{tri}}{\Phi_0},\\
    \phi_{10} &= \phi_1,\\
    \phi_{20} &= \phi_2.
\end{align}

In addition to the global phase frustration, the magnetic field suppresses
the critical current of each individual weak link.
We account for this phenomenologically by using a Fraunhofer dependence to the amplitude $I_{c,ij}$,
\begin{equation}
I_{\mathrm{c},ij}(B)
=
I_{\mathrm{c},ij}(0)
\frac{
\sin\left(
\pi B A_{k}^{\mathrm{eff}}/\Phi_0
\right)
}{
\pi B A_{k}^{\mathrm{eff}}/\Phi_0
},
\end{equation}
where $A_{k}^{\mathrm{eff}}$ denotes the effective area of the junction between terminals $i$ and $j$.
The Josephson contributions to the terminal currents then become
\begin{align}
I_{\mathrm{J},1}
&=
I_{\mathrm{c},12}(B)
\sin\left(
\phi_1-\phi_2+2\pi\frac{\Phi_\mathrm{tri}}{\Phi_0}
\right)
+
I_{\mathrm{c},10}(B)\sin\phi_1,
\\
I_{\mathrm{J},2}
&=
-
I_{\mathrm{c},12}(B)
\sin\left(
\phi_1-\phi_2+2\pi\frac{\Phi_\mathrm{tri}}{\Phi_0}
\right)
+
I_{\mathrm{c},20}(B)\sin\phi_2.
\end{align}
These equations are then fed into the solver of the RCSJ model published by Gupta et al. in Ref. \cite{Gupta2023}. To better identify the exact shape of the CCC, we plot the average resistance of both terminals to ground, which allows for clear recognition of all relevant features.
The simulated reduced-resistance arms appear somewhat wider than in the measured data, which we attribute mostly to the lack of Joule heating in  the model.  
The spatially continuous phase distribution around the device is represented by a single global phase-winding parameter, which is an approximation valid for small magnetic fields.

\bibliography{bib_sup}
